\documentstyle[12pt]{article}
\begin{document}
\def\Tm{T^m}
\def\tm{t_m}
\def\Tdm{T^m_d}
\def\tdm{t_m^d}
\def\ot{\otimes}
\def\br{\bf R}
\def\al{\alpha}
\def\bt{\beta}
\def\th{\theta}
\def\ga{\gamma}
\def\vth{\vartheta}
\def\da{\downarrow}
\def\ua{\uparrow}
\def\de{\delta}
\def\lm{\lambda}
\def\b{\beta}
\def\k{\kappa}
\def\om{\omega}
\def\az{\vert z\vert}
\def\Om{\Omega}
\def\si{\sigma}
\def\w{\wedge}
\def\od{\sqrt{L^2+1}}
\def\Tn{T^n}
\def\tn{t_n}
\def\Tdn{T^n_d}
\def\tdn{t_n^d}
\def\e{\varepsilon}
\def\ti{\tilde}
\def\js{{1\over 4}}
\def\I{{\cal I}}
\def\L{{\cal L}}
\def\D{{\cal D}}
\def\G{{\cal G}}
\def\bz{{\bar z}}
\def\E{{\cal E}}
\def\B{{\cal B}}
\def\M{{\cal M}}
\def\K{{\cal K}}
\def\J{{\cal J}}
\def\tJ{\ti{\cal J}}
\def\R{{\cal R}}
\def\d{\partial}
\def\la{\langle}
\def\ra{\rangle}
\def\bc{{\bf C}}
\def\st{\stackrel{\w}{,}}
\def\lta{\leftarrow}
\def\rta{\rightarrow}
\def\scu{$SL(2,\bc)/SU(2)$~ $WZW$  }

\def\be{\begin{equation}}
\def\ee{\end{equation}}
\def\jp{{1\over 2}}

\def\tD{\Delta^*}
\def\slc{SL(2,{\bf C})}
\begin{titlepage}
\begin{flushright}
{}~
IML 02-28\\
hep-th/0211177
\end{flushright}

\vspace{1cm}
\begin{center}
{\Huge \bf  $D$-branes in the Euclidean $AdS_3$ and $T$-duality}\\
[50pt]{\small
{\bf Ctirad Klim\v{c}\'{\i}k}
\\ ~~\\Institute de math\'ematiques de Luminy,
 \\163, Avenue de Luminy, 13288 Marseille, France}
\end{center}

\vspace{0.5 cm}
\centerline{\bf Abstract}
\vspace{0.5 cm}

  We show that
$D$-branes in the Euclidean $AdS_3$ can be
naturally associated to the maximally isotropic
subgroups of the Lu-Weinstein double of $SU(2)$. This picture
makes very transparent the residual loop group symmetry
of the $D$-brane configurations and
gives also immediately  the   $D$-branes shapes and
 the $\sigma$-model
boundary conditions  in the   de Sitter $T$-dual of
the $SL(2,C)/SU(2)$ WZW model.

\end{titlepage}
\section{Introduction}
An even-dimensional Lie group $D$ equipped with  a maximally Lorentzian
biinvariant metric is called a   Drinfeld
 double\footnote{Here we commit a little  abuse of terminology.
 In fact,
for the  standartly defined  Drinfeld double \cite{D},
it must be moreover true  that
$Lie(D)$ is   the
 (vector space) direct
sum $Lie(G_1) +Lie(G_2)$. The latter condition was even   present
 in the original version of the Poisson-Lie $T$-duality
\cite{KS1}. However, as it  was shown later in \cite{KS2},
 it can be released and
 the duality  continues to take place.}
  \cite{D,KS2} if it has at least two
maximally isotropic  Lie subgroups $G_1$ and $G_2$, not related
by an inner automorphism   in $D$. It was shown
in \cite{KS2}, that the Drinfeld double
$D$
and  an unipotent linear operator $\E$ on its Lie algebra $\D$
naturally define mutually dual closed string  $\si$-models
on the targets $D/G_1$ and $D/G_2$, respectively.
By considering moreover an element $d\in D$ and another maximally
 isotropic
subgroup $M$ of $D$, the quadruple $(D,\E,d,M)$ defines
a mutually dual pair of open string $\si$-models \cite{KS3,KS4}.
 In particular, the  $D$-brane submanifolds of the targets $D/G_1$
and $D/G_2$ are, respectively,
  the coset projections
$\pi_{G_1}$ and $\pi_{G_2}$
of $Md\subset D$ to $D/G_1$ and $D/G_2$.

 We have recently shown \cite{K},
that for $D=SL(2,\bc)$,
$G_1=SU(2)$, $G_2=SL(2,\br)$ and an appropriate $\E$, the
corresponding
 bulk $\si$-model on $D/G_1$
is nothing but  the $SL(2,{\bc })/SU(2)$ WZW  model \cite{Gaw}
describing   strings
in the Euclidean $AdS_3$, while the dual model living on $D/G_2$
captures the string dynamics in the three-dimensional de Sitter
space.
 In this paper, we shall enlarge our discussion to incorporate
the open string $\si$-models. We shall show, in particular, that
all $AdS_3$ $D$-brane boundary conditions, recently considered by
Ponsot, Schomerus and Teschner \cite{PST}, can be obtained
by choosing appropriately the data $d$ and $M$ of the general
construction
indicated above. This  means that we will be able to write down
 immediately
also  de Sitter duals of the $AdS_3$ $D$-branes.

From the technical point of view, this paper just makes explicit
the  general theory of D-branes T-duality \cite{KS3,KS4}  for the
particular case
of the $\scu$ model. We believe that it is worth working out
this example for  two reasons:
1) the $\scu$ model plays an important  role in the AdS/CFT
 correspondance
therefore its duality structure should be of interest; 2) the model
is conformal and
has a loop group symmetry hence it is so far the best candidate for
testing the quantum status of the Poisson-Lie T-duality.

The Poisson-Lie T-duality is relevant for the $\scu$ model
because the latter is Poisson-Lie symmetric \cite{K}. It is less
obvious, however, why the D-branes boundary conditions considered
by Ponsot, Schomerus and Teschner \cite{PST}  fits in the
framework of the Poisson-Lie T-duality.  In fact, the authors
of \cite{PST}
 were guided by the requirement of the
residual loop group symmetry of the open string $\si$-model. This
condition gave them the shapes of the $AdS_3$ $D$-branes as well as
$B$-fields encoding the boundary condition. On the other
hand, the authors of \cite{KS3,KS4}
were using the criterion of T-dualizability for finding the shape of
$D$-branes
and boundary conditions on them. Remarkably, those very differently
looking criterions give the same $D$-branes for the $\scu$ model.
We shall offer the explanation   of this fact
based on the  association of the $D$-branes boundary
conditions to the maximally isotropic subgroups of the Drinfeld
 double.

In section 2, we review the general concept of the open string
$\si$-models with particular emphasis on the global characterization
of the $D$-branes boundary conditions.  Then we review the
Poisson-Lie T-duality in the presence of $D$-branes.
In section 3, we work out in detail the case of the  $\scu$ model.
We consider
the maximally isotropic subgroups of the Drinfeld double $SL(2,\bc)$,
we derive from them the $AdS_3$ $D$-branes studied by Ponsot,
Schomerus and Teschner and
 we describe the de Sitter
duals of the $AdS_3$ $D$-branes.
 Finally,  in section 4, we explain the
existence of their residual loop group symmetry.

\section{Open strings and $D$-branes}
\subsection{Generalities}

\noindent 1.
In the closed string case, the classical dynamics of  nonlinear
$\si$-model is completely characterized by
a metric $d\sigma^2=\jp G_{ij}dx^idx^j$
 and by a closed three-form  $H$  on a target
manifold $T$. When  $H=dB$ for some two-form $B=
\jp B_{ij}dx^i\w dx^j$,
 the
latter is called the $B$-field and the (Euclidean) action of the
$\si$-model can be written as
$$S[x]=i\int d\bz\w dz (G_{ij}(x)+B_{ij})\d_\bz x^i\d_z x^j.$$
  But also when $H$ is cohomologically
nontrivial (i.e. there is no globally defined potential $B$) the
classical model can
be perfectly defined. Indeed, consider for definiteness the Riemann
sphere as the closed string world-sheet. Then the (Euclidean)
$\si$-model
action can be cast in the WZW-like way:
$$S[x]=i\int_{\d\Omega} d\bz\w dz G_{ij}(x) \d_\bz x^i\d_z x^j+
i\int_\Om
 {\ti x}^*H.\eqno(1)$$
Here $\Om$ is a three-dimensional ball whose boundary
$\d\Om$ is the Riemann sphere, the notation $\ti x^*H$
means the pull-back of $H$ to $\Om$ via $\ti x$,
and the latter   is the (extension) map from the ball $\Om$
to the target $T$ whose boundary value $\ti x^i\vert_{\d \Om}$
is the
$\si$-model configuration $x^i(z,\bz)$.   Of course, the classical
action (1) can be defined only when
  the topology of the target $T$ permits to extend every map
$x$ on $\d \Om$ to the map $\ti x$ on $\Om$. Moreover, this
extension need
not be unique. Two different extensions
can give a different value of the action $S$.
Although   the
 classical dynamics (i.e. the field equations) is well defined
and does  not depend on this
ambiguity, the path integral quantization is impossible unless
the ambiguity of the action is $2\pi in$ with $n$ an integer.
If this is the case, the three-form $H$   defines  an integer-valued
cocycle in the singular cohomology  of $T$.

\vskip1Pc
\noindent 2.
The first complete discussion (covering also the case of the
non-exact $H$)
of the open string case was given in \cite{KS4}.  The $\si$-model
is again characterized by the target $T$,
the metric $d\si^2$ and the closed three-form field
$H$  but also  by a two-form $\al$
living on a submanifold  $P$ of $T$. Needless to say, the
submanifold
$P$ is referred to as a $D$-brane.
 The boundary of the world-sheet has to lie
in $P$ and the restriction $H_P$ of $H$ on $P$ must admit $\al$ as
its potential,
i.e. $d\al=H_P$.

It is instructive to write down the open string variational
principle in the
particular case where the   world-sheet
is the disc $\vert z\vert\leq 1$. We shall view this disc as the
southern half of the Riemann sphere and denote it as $S_\da$.
We shall need to denote also the northern hemisphere as $S^\ua$.
Let now $x^i(z,\bz)$ be a $\si$-model configuration   defined on
$S_\da$ and having the boundary values  $x^i(\vert z\vert =1)$
in the
brane $P\subset T$. The action of this open string
configuration is then given by
$$S[x]=i\int_{S_\da} d\bz\w dz G_{ij}(x) \d_\bz x^i\d_z x^j+
i\int_\Om
 {\ti x}^*H -i\int_{S^\ua}x_P^*\al.\eqno(2)$$
Here $x_P^i$ is an arbitrary map from $S^\ua$ to the brane $P$
coinciding with the open string configuration $x^i(z,\bz)$ on the
common boundary of $S^\ua$ and $S_\da$. In other words:
$x^i(\vert z\vert =1)=x_P^i(\vert z\vert =1)$. The (extension) map
$\ti x^i$ from $\Om$ to $T$ must now fulfil the following boundary
conditions:
$\ti x^i\vert_{S_\da}=x^i$ and $\ti x^i\vert_{S^\ua}=x_P^i$.

We note that the open string action principle involves two
 extensions:
first we extend the $\si$-model configuration $x^i(z,\bz)$ from
the hemisphere $S_\da$ to the whole sphere $\d\Om=S_\da\cup S^\ua$
 by choosing
 $x_P^i$
on $S^\ua$ (in a sense we complete the sphere in the brane $P$)
and then we extend $x^i\cup x_P^i$ to the map $\ti x^i$.
The existence
and ambiguity of these extensions depend on the topology of $T$
and $P$.
The detailed discussion of this issue is given in \cite{KS4}. Here
we only note three things: 1)  for all choices of $T$ and $P$
 considered in this paper, the
existence of the extensions will be guaranteed; 2) the existence
of the extensions itself guarantees
 the  unambiguous definition of the field equations
of the model (essentially due to the property $d\al=H_P$); 3)
the non-unicity  of the extensions is an important issue for the
  {\it quantization}. Actually,
 the path integral quantization is impossible unless
the ambiguity of the action is $2\pi in$ with $n$ an integer.
If this is the case, the pair  $(H,\al)$   defines an
integer-valued
cocycle in the relative singular
 cohomology  of $T$ with respect to $P$ \cite{KS4}.

The reader might not be accustomed with writing the
non-metric part of the open string $\si$-model (2) as
$$i\int_\Om \ti x^* H-
i\int_{S^\ua}{x_P^*} \al .\eqno(3)$$ Usually the people write
 instead
$$i\int_{S_\da}x^*B+i\int_{\vert z\vert =1} x^*A,\eqno(4)$$
where $B$ is the  two-form potential $B$ such that $dB=H$ and
 $A$ is a one-form
on the brane $P$. The advantage of the formulation (4) is clear:
we do not need to consider any extension $\ti x$  or $x_P$
 whatsoever.
However, the formulation (4) is less general since it can be
derived from (3) only when two requirements are satisfied: 1) the
three-form $H$ must be exact on the target $T$
(i.e. $H=dB$ for some two-form $B$ on $T$); 2) the (closed)
 two-form $(B_P-\al)$ must be exact on the brane $P$
(i.e. $B_P-\al=dA$ for some one-form $A$ on $P$).
Of course, here $B_P$ is the restriction of the form $B$ on the
 brane $P$.
It is immediate to verify that (3) yields (4)
if these two conditions are satisfied.

It turns out  that even in the cases
when one can reduce the description (3)  to (4), it is
sometimes better to work with the invariant description (3).
For example, the symmetry structure of the brane configuration
is typically more transparent  in the invariant formulation (3).

\vskip1Pc
\noindent 3. Consider now a closed string background
$(T,d\si^2, H)$.
As we already know, the open string generalization can be
defined, if we add
two more data $(P,\al)$. Of course,
there are many possible quintuples
 $(T,ds^2,H,P,\al)$ to consider.  We may wish
to impose further  restrictive conditions on these data,
according to the aspects of the
open string dynamics that we wish to study.
There are two examples of such additional conditions:

\vskip1pc
\noindent  i) In the framework of the WZW-like models, we wish that
the part of the Kac-Moody symmetry be preserved by the boundary
conditions. Such requirement typically ensures the preservation
of the conformal symmetry  of the open string model.

\noindent ii) If the closed string model $(T,d\si^2,H)$
admits a T-dual
$(T',d\si'^2,H')$,
we require that also the open string model $(T,d\si^2,H,P,\al)$
 admits
a T-dual $(T',d\si'^2,H',P',\al')$.
\vskip1pc
\noindent Ponsot, Schomerus and Teschner
\cite{PST} have looked for the $D$-brane boundary conditions
satisfying
the criterion i). In distinction to it,
we shall organize our paper from the point of view of the
requirement ii). In fact,
if the bulk $(T,d\si^2,H)$-model is just the Poisson-Lie
dualizable $\si$-model  on the target $D/G_1$ (see Introduction),
 then
there is a simple method \cite{KS3,KS4} to generate the D-brane
shapes $P$
and the
boundary  conditions $\al$
   fulfilling
the requirement ii). In fact, take any element $d$ of the
corresponding
Drinfeld double $D$   and any
   maximally isotropic
subgroup $M$ of   $D$. We shall see in a while
that the pair $(M,d)$ then completely determines
the T-dualizable quintuple $(T,d\si^2,H,P,\al)$.
In the case of the $\scu$ model,   there is another
 pleasing circumstance of this construction; namely,
 the obtained T-dualizable quintuple $(T,d\si^2,H,P,\al)$
 preserves the residual
$M$-loop group symmetry hence it
automatically satisfies also
the criterion i).  This observation explains, why we obtain
the same brane configurations in the Euclidean $AdS_3$ as Ponsot,
Schomerus and Teschner  \cite{PST}.

\subsection{D-branes and Poisson-Lie T-duality}
\noindent 1.
This subsection is a brief review of the results\footnote{Some
conventions in this paper are changed with respect to
our recent article \cite{K}. For example, the relation between
the world-sheet coordinates $\si,\tau$ and $z,\bz$, the
normalization of the line element $d\si^2$ etc. We did those
changes
in order to have the same conventions as the paper \cite{PST}
of Ponsot, Schomerus and Teschner. This will permit us the
direct
comparison of our results with theirs.} \cite{KS3,KS4,K}.
It serves to keep complet the logical skeleton of the paper. However,
the reader
wishing to enter technical details must  consult those
papers.
Consider the following first order  action   on the Drinfeld
double $D$:
$$S=\jp\int_{\d\Om} d\si\w  d\tau (\d_\tau ll^{-1},
\d_\si ll^{-1})_\D
+{1\over 12}\int_\Om(dll^{-1}\st [dll^{-1}\st dll^{-1}])_\D$$
$$-\jp\int_{\d\Om} d\si \w d\tau (\d_\si ll^{-1},\E
\d_\si ll^{-1})_\D.\eqno(5)$$
Here $l(\si,\tau)$ is a map from the world-sheet into the double
and the world-sheet coordinates are defined as
$$z=\tau+i\si,\quad \bz =\tau-i\si,\quad \d_z=\jp(\d_\tau-i\d_\si),
\quad \d_\bz=\jp(\d_\tau +i\d_\si).$$
The operator $\E:Lie(D)\to Lie(D)$ is self-adjoint with respect
to the (metric) bilinear form $(.,.)_\D$ on $\D=Lie(D)$ and
it holds $\E^2=\pm Id$. Actually, here we shall consider
only the case $\E^2=-Id$, because it leads to the T-duality
between $\si$-models with real Euclidean actions \cite{K}.

For the moment,
 the action (5) is considered on a world-sheet without boundaries
and
it encodes the closed strings Poisson-Lie T-duality between the
 targets
$T_1=D/G_1$ and $T_2=D/G_2$.   Indeed,
consider   the coset $D/G_j$ and parametrize\footnote{If there
exists no global
section of this fibration, we can choose several local sections
covering
the whole
base space  $D/G_j$.} it by the
elements $f_j$ of $D$. With this parametrization of $D/G_j$, we
 may parametrize
the surface  $l(\tau,\si)$ in the double as follows
$$l(\tau,\si)=f_j(\tau,\si)g_j(\tau,\si),\quad  g_j\in G_j\eqno(6)$$
and there is no summing over $j$.
 The action (5) then becomes
$$S=\jp\int_{\d\Om} (f_j^{-1}\d_\tau f_j,f_j^{-1}\d_\si f_j)_\D +
{1\over 12}\int_\Om(df_jf_j^{-1}\st
 [df_jf_j^{-1}\st df_jf_j^{-1}])_\D+$$
$$+\int_{\d\Om} (\d_\si g_jg_j^{-1},f_j^{-1}\d_\tau f_j)_\D-
\jp\int_{\d\Om}
(f_j^{-1}\d_\si f_j  +\d_\si g_jg_j^{-1},\E_{f_j}(f_j^{-1}\d_\si f_j +
\d_\si g_jg_j^{-1}))_\D,\eqno(7)$$
where $\E_{f_j}=Ad_{f_j^{-1}}\E Ad_{f_j}$   and we tacitly
suppose the measure
$d\si \w d\tau$ present in the formula. Now we note that the
expression (7)
is Gaussian in the $Lie(G_j)$-valued variable
$ \d_\si g_jg_j^{-1}$.  The most useful strategy to solve it away
is to pick
up some basis $S_j^a$ in $Lie(G_j)$, write $\d_\si g_jg_j^{-1}=
\mu_{ja}S_j^a$
and integrate
away $\mu_{ja}$. This gives
$$S=\jp\int_{\d\Om} d\bz\w dz( \d_\bz f_jf_j^{-1},
\d_z f_jf_j^{-1})_\D+
{1\over 12}\int_{\Om} (df_jf_j^{-1}\st [df_jf_j^{-1}\st
df_jf_j^{-1}])_\D+$$
 $$+i\int_{\d\Om} d\bz \w dz(f_j^{-1}\d_\bz f_j,S_j^a+i
\E_{f_j} S_j^a)_\D
(A_{f_j}^{-1})_{ab}(S_j^b,f_j^{-1}\d_z
f_j)_\D.\eqno(8)$$
where
$$A_{f_j}^{ab}=(S_j^a,\E_{f_j} S_j^b)_\D.\eqno(9)$$
We note that in spite of the explicit presence of the imaginary unit
in this formula,
the $\si$-model action (8) is always real. The duality is the
equivalence of the models (8) for different $j$.

By the way,
  the  $\si$-model like (8) can be
associated to every maximally isotropic subgroup  of $\D$
provided the
corresponding
matrix $A_f$ is invertible. The target of such a $\si$-model
is the coset
of the Drinfeld double by this subgroup.
\vskip1Pc
 \noindent 2. We wish to consider the first order action (5)
on the hemisphere (or disc) $S_\da$.
For this, we choose some maximally
isotropic\footnote{This means that the
$Lie(M)$ is the maximally isotropic subspace of $Lie(D)$
with respect
to the canonical indefinite metric $(.,.)_\D$.} subgroup $M$
of $D$ and
an element $d\in D$ and require that $l(z,\bz)\in Md$ for
$\vert z\vert =1$.
The
 submanifold
$Md$ is the left $M$-orbit of the element $d$ and we shall call
it a
first order
$D$-brane. The first
and the third terms of the action (8) can be perfectly
defined as integrals
over $S_\da$ only.
 As it is discussed
in \cite{KS4} and in Section 2.1, giving sense to the second
term requires
to choose a two-form
$\al_D$   on the first order  brane $Md$ such that $d\al_D$
is the restriction
of the three-form $1/6(dll^{-1}\st[dll^{-1}\st dll^{-1}])_\D$
on $Md$.
In the spirit of Section 2.1,
the open string first order  action then reads:
$$S=\jp\int_{S_\da} d\si\w  d\tau (\d_\tau ll^{-1},
\d_\si ll^{-1})_\D
-\jp\int_{S_\da} d\si \w d\tau (\d_\si ll^{-1},
\E\d_\si ll^{-1})_\D.$$
$$+{1\over 12}\int_\Om(dll^{-1}\st [d l l^{-1}\st dll^{-1}])_\D
-\jp\int_{S^\ua}l^*\al_D.\eqno(10)$$
We ask the reader to excuse us some abuse of  notation: we should
have denoted by $l$ the true $S_\da$-open string configuration,
by $l_{Md}$
its $S^\ua$ extension in the first order brane $Md$
and by $\ti l$
the extension of both $l$ and $l_{Md}$ to the ball $\Om$.
Instead, we have used everywhere $l$ and hoped not to
cause a confusion.

 Now remark,
that the $D$-invariance  of the metric $(.,.)_\D$ means
that not only $M$ but
also $Md$ is the isotropic surface  in the
double $D$. This fact implies that
the restriction of
$1/6(dll^{-1}\st [dll^{-1}\st dll^{-1}])_\D$ on $Md$
simply vanishes.
Thus we  fully define the open string model (10)  by
choosing a closed
form $\al_D$
on $Md$. We make the simplest possible choice and set $$\al_D=0.$$
Let $G_1$ and $G_2$ be two maximally  isotropic subgroups of $D$.
 If the
respective matrices $A_f$  given by (9) are invertible for both
choices of $G_1$ and  $G_2$
we know that there is the bulk  $T$-duality between the model
(8) on $D/G_1$
and its  counterpart  on $D/G_2$. This duality takes place
also in the
open string case if we start with the action (10),
the boundary condition
$l(\vert z\vert =1)\in Md$ and $\al_D=0$.

We shall not repeat here the derivation of the  mutually
dual quintuples $(T_j,d\si_j^2,H_j,P_j,\al_j)$,
$j=1,2$ obtained from the first-order action (10). It
has been done in \cite{KS4} at it is based on the following
variant of the Polyakov-Wiegmann formula:
$$(fg)^*WZW(l)=f^*WZW(l) +g^*WZW(l) -d(f^*(l^{-1}dl)\st
 g^*(dll^{-1}))_\D.
\eqno(11)$$
Here
$$WZW(l)\equiv 1/6(dll^{-1}\st [dll^{-1}\st dll^{-1}])_\D$$
and the two maps from $D$ to $D$: $f(l)=f$ and $g(l)=g$ are
induced by the
decomposition (6).  As usual, $*$ denotes the pull-back
of the forms under the mappings to the group manifold
$D$. The result of the derivation is as follows:

\vskip1pc
\noindent 1) The manifold $T_j$ is the coset $D/G_j$.

\noindent 2-3) The metric $d\si_j^2$  and the $H_j$-field
are given by the bulk $\si$-model (8). Explicitely:
 $$ds_j^2= \jp (f_j^{-1}df_j,
S_j^a)_\D(A_{f_j}^{-1})_{ab}(S_j^b,f_j^{-1}df_j)_\D;\eqno(12)$$
$$H_j={i\over 2}d\{(f_j^{-1}df_j,\E_{f_j}S_j^a)_\D\w
(A_{f_j}^{-1})_{ab} (S_j^b,
f_j^{-1}df_j)_\D\}-
{i\over 12} (df_jf_j^{-1}\st [df_jf_j^{-1}\st df_jf_j^{-1}])_\D.$$

\noindent 4) The D-brane $P_j$   in $D/G_j$  is the
image of $Md\subset D$ under the coset projection
$\pi_{G_j}:D\to D/G_j$.

\noindent 5) The form $\al_j$ on $P_j$ is given
by
$$\al_j ={i\over 2}(f_j^{-1}df_j,\E_{f_j}S_j^a)_\D\w
(A_{f_j}^{-1})_{ab} (S_j^b,
f_j^{-1}df_j)_\D -{i\over 2}(f_j^{-1}df_j\st
dg_j(f_j)g_j(f_j)^{-1})_\D.
\eqno(13)$$
Recall that $f_j\in D$ are the representatives of the
coset elements in $D/G_j$ and   $g_j(f_j)$ is any map
from $P_j$ to $G_j$, such
that $f_jg_j(f_j)\in Md$. The ambiguity in the definition of the
map $g_j:P_j\to G_j$ does not influence the form $\al_j$.
 Moreover,
the maps $g_j$  may exist only locally on $P_j$. However,
the form $\al_j$ is defined {\it globally} on $P_j$ and has
to be glued
from the maps $g_j$ defined on local charts forming a
covering of $P_j$.

\section{D-branes in the Euclidean $AdS_3$ }
\subsection{The bulk story}
Here we apply the  general  results of Section  2
to the $SL(2,\bc)/SU(2)$ WZW model.
The Drinfeld double $D$ is the   group $SL(2,\bc)$ viewed
as a real group
and the biinvariant maximally Lorentzian metric on it is
naturally induced from the following non-degenerate invariant
symmetric  bilinear form  $(.,.)_\D$  on its Lie algebra
 $\D=sl(2,\bc)$:
 $$(x,y)_\D={\rm ImTr}(xy).$$
In other words, the indefinite metric is given by the
imaginary part of the trace in the fundamental representation
of $sl(2,\bc)$.
For the maximally isotropic subgroups $G_1$ and $G_2$ we take,
 respectively,
$SU(2)$ and $SL^a(2,\br)$. The superscript
$a$ means that the group $SL(2,\br)$ is atypically embedded
into $SL(2,\bc)$ according to the following formula \cite{K}:
$$\left(\matrix{\mu & i\nu\cr i\rho &\lm}\right)
\in SL^a(2,R),\quad
 \mu,\nu,\rho,\lm\in\br,\quad \mu\lm +\nu\rho=1.$$
 In fact, our embedding  $SL^a(2,\br)$ is conjugated
to the standard one (real matrices with unit determinant)
by the following
element of
$SL(2,\bc)$  :
$${1\over  2}\left(\matrix{1+i & 1+i\cr i-1 &1-i}\right).$$
We recall that $\si$-models (8) on $D/G$ and on $D/G'$ have the same
target geometry
if $G$ and $G'$ are conjugated in $D$. This means that we could
equally well consider the standardly embedded $SL(2,\br)$ as the
maximally isotropic subgroup $G_2$. Our atypical choice
 $G_2=SL^a(2,\br)$
is motivated
by an effort to make as straigthforward as possible
the comparison of our results with those
of Ponsot, Schomerus and Teschner \cite{PST}.
We also note that the isotropy of the Lie algebras
$su(2)$ and $sl^a(2,\br)$
is clear because they are both the {\it real} forms of $SL(2,\bc)$.

Finally, we need
the operator $\E$. It is simply the multiplication by the imaginary
unit in $sl(2,\bc)$ viewed as the {\it real} Lie algebra.
Now we can write  the  first order action (10) in this
particular case:
$$S=\jp\int_{\d\Om} d\si\w  d\tau {\rm ImTr}
(\d_\tau ll^{-1}\d_\si ll^{-1})
+{1\over 12}\int_\Om{\rm ImTr}(dll^{-1}\w [dll^{-1}\st dll^{-1}])$$
$$-\jp\int_{\d\Om} d\si \w d\tau{\rm ImTr}
 (\d_\si ll^{-1} i\d_\si ll^{-1}).
\eqno(14)$$
This expression can be cast even more simply as
$$S=\int_{\d\Om}d\si\w  dz {\rm ImTr}(\d_z ll^{-1}\d_\si ll^{-1})
+{1\over 12}\int_\Om{\rm ImTr}(dll^{-1}\w [dll^{-1}\st dll^{-1}]). $$
 We recall
that the  T-duality relates the $\si$-models living on
$SL(2,\bc)/SU(2)$
and on $SL(2,\bc)/SL^a(2,\br)$.
Since we already know the general formulae (12)
and (13), we need only
to decompose the world-sheet $l(\si,\tau)$ in the
double $D=SL(2,\bc)$
as $l=f_jg_j$, $j=1,2$ (cf. (6)). The section  $f_2$
was constructed
in \cite{K}:
$$f_2=\left(\matrix{\cos{\vth}+
i{L \over \sqrt{L^2+1}} \sin{\vth} & i{1\over \od}\sin{\vth}
 \cr i{1\over \od}\sin{\vth} &\cos{\vth}
-i{L \over \sqrt{L^2+1}}\sin{\vth}} \right)
\left(\matrix{\cos{\chi} & \sin{\chi}\cr -\sin{\chi} &
 \cos{\chi}}\right)\left(\matrix{1 & L \cr 0 &1}\right),\eqno(15)$$
where $0\leq \vth\leq \pi$, $0\leq \chi\leq \pi/2$ and  $L\in\br$.

For the case  $j=1$, we  parametrize the coset  $SL(2,\bc)/SU(2)$
differently as in \cite{K}. For the sake of being
notationally as close
as possible to the paper \cite{PST}, we take for $f_1$
the following
section
$$f_1=\left(\matrix{e^{\phi/2}&0\cr
\ga e^{\phi/2} &e^{-\phi/2}}\right),\eqno(16)$$
where $\phi\in\br$ and $\ga\in\bc$.
Now we can directly calculate from (12) the $\si$-model metrics and
$H$-fields for the both cases $j=1,2$:
$$d\si_1^2= -\js(d\phi^2+e^{2\phi}d\ga d\bar\ga),\quad
H_1=-\jp e^{2\phi} d\phi\w d\bar\ga \w d\ga;\eqno(17)$$
$$d\si_2^2=-\js {(dL)^2\over L^2+1}\sin^2{2\chi}\sin^2{2\vth}+
(L^2+1)(d\chi^2+\sin^2{2\chi}d\vth^2)+$$
$$+\cos{2\vth}dL d\chi
-\jp\sin{4\chi}\sin{2\vth}dLd\vth; $$
 $$H_2=-4i\od \sin{2\chi} dL\w d\chi\w d\vth .\eqno(18)$$
$d\si_1^2$ and $H_1$ are equal  to the metric and $H$-field of the
$\scu$ model \cite{Gaw,PST}. The metric $d\si_2^2$ turns out
to be the de Sitter
metric written in appropriate coordinates (cf.\cite{K}). $d\si_2^2$
 and $H_2$ together define
the de Sitter $\si$-model introduced in \cite{K}.

It is
  very convenient to
   introduce the following Hermitian matrices
$$h=f_1f_1^\dagger,\quad s=f_2\si_1f_2^\dagger,\eqno(19)$$
where $\si_1=\left(\matrix{0&1\cr 1&0}\right)$ is the
Pauli matrix and $\dagger$ means the Hermitian conjugation
of matrices.  In terms of $h$ and $s$ the $\si$-model
backgrounds (17) and
(18), can be respectively rewritten as
$$d\si_1^2 =-{1\over 8}{\rm Tr} (dhh^{-1} dh h^{-1}),\eqno(20a)$$
$$H_1=-{1\over 24}{\rm Tr}(dhh^{-1}\w [dhh^{-1}\st dhh^{-1}]).
\eqno(20b)$$
$$d\si_2^2 =-{1\over 8}{\rm Tr} (dss^{-1} dss^{-1}),\eqno(21a)$$
$$H_2=-{1\over 24}{\rm Tr}(dss^{-1}\w [dss^{-1}\st dss^{-1}]).
\eqno(21b)$$
 In other words, the closed string $\scu$ action   can be written
as
$$S=-{i\over 4}\int_{\d\Om} d\bz\w dz {\rm Tr}(\d_\bz hh^{-1}
\d_z hh^{-1})
-{i\over 24}\int_\Om{\rm Tr}(d\ti h\ti h^{-1}
\w[d\ti h\ti h^{-1}\st d\ti h\ti h^{-1}])\eqno(22)$$
and the closed string de Sitter action as
$$S=-{i\over 4}\int_{\d\Om} d\bz\w dz {\rm Tr}(\d_\bz ss^{-1}
\d_z ss^{-1})
-{i\over 24}\int_\Om{\rm Tr}(d\ti s\ti s^{-1}\w
[d\ti s\ti s^{-1}\st d\ti s\ti s^{-1}]).\eqno(23)$$
 Note that in (20) - (23) there is the full trace, not only its
imaginary part.  The original and the dual model look pretty
 the same
but we must realize that the determinant of $s$ is $(-1)$ and
 that of $h$
is $1$. Moreover, the trace of $h$ is positive.

The T-duality for the open strings depends on the choice of the
first order $D$-branes $Md$. We shall separately consider three
cases:   1) $M=SL^a(2,\br)$; 2) $M= SU(2)$; 3)  $M=AN$.

\subsection{$AdS_2$ branes}
We start with the case $M=SL^a(2,\br)$.
The possible first order $D$-branes are then the submanifolds $Md$
of $SL(2,\bc)$
with $d$ being a fixed element of $SL(2,\bc)$. We do not make here
 the most
general choice of $d$; instead, we consider only $d$ having form
$$d=\left(\matrix{1 & 0\cr\jp c &1}\right),\quad c\in\br.$$
The motivation for this particular choice is simple:
when we switch from the first order description  (10) to the
second order one (22), it  reproduces
the $AdS_2$ branes considered in $\cite{PST}$.
\vskip1pc
\noindent Thus our first order $D$-branes are three-dimensional
submanifolds of $SL(2,\bc)$
of the form
$$p=Md=SL^a(2,{\bf R})d=\left(\matrix{\mu& i\nu\cr i\rho & \lm}
\right)
\left(\matrix{1 & 0\cr\jp c&1}\right)=
\left(\matrix{\mu  +\jp{i\nu c }& i\nu\cr i\rho+
\jp{\lm c} & \lm}\right),\eqno(24)$$
where the real number $c$ is fixed and
$\mu,\nu,\rho,\lm\in\br$ vary while respecting
the constraint $\mu\lm +\nu\rho=1$.
\subsubsection{The target $D/G_1=SL(2,\bc)/SU(2)$}
Now we want to find the shape of the  corresponding
 $D$-brane $P_1$
in the target $D/G_1=SL(2,\bc)/SU(2)$.
As it is well-known, the coset $SL(2,\bc)/SU(2)$ can be
identified with
the group
$AN$, whose elements are the matrices of the form (16).
  There is another natural way to view this coset, namely, as a set
of Hermitean matrices in $\slc$ with a positive trace and unit
determinant. They  can be  parametrized as (cf. \cite{PST})
$$ h=f_1f_1^\dagger=\left(\matrix{e^\phi &  e^\phi \bar \ga\cr
e^\phi\gamma &e^\phi\ga\bar\ga
+e^{-\phi}}\right), \quad  \phi\in\br,\  \ga\in\bc.\eqno(25)$$
The canonical coset projection map $h:\slc\to \slc/SU(2)$ is
simply given by
$$h(l)=ll^\dagger,\quad l\in\slc.\eqno(26)$$
 It turns out that
the restriction of the map $h$ to the subgroup $AN$ gives a
 diffeomorphism
between the subgroup  $AN$ of $SL(2,\bc)$
and the space (25).
Now we are ready to find the shape of the $D$-brane $P_1$
in $D/G_1$.
It is given by the coset projection of the first order
$D$-brane (24),
or, in other words,
  by the Hermitian matrices of the form
$$h(p)=\left(\matrix{\nu^2(1+\js c^2)+
\mu^2 &   -i\mu\rho+i\nu\lm(1+\js c^2)+\jp c
\cr   +i\mu\rho-i\nu\lm(1+\js c^2)+\jp c  &
\lm^2(1+\js c^2)+\rho^2 }\right).$$
Comparing with the parametrization (25), we observe that
the $D$-brane
$P_1$ in the coset $\slc/SU(2)$ is characterized by the equation
$$e^\phi(\ga+\bar\ga)=c\eqno(27)$$
with a constant $c$. This is exactly the $AdS_2$ brane
considered in \cite{PST}.

As we have said, the first order action (10) together with
the choice of the
first order $D$-brane (24) (and $\al_D=0$)
determines the open string quintuple
$(T_1,ds_1^2,H_1,P_1,\al_1)$.
So far we have determined  four its elements: the target
$T_1=\slc/SU(2)$ is parametrized as in (25), then
$$ds_1^2=-\js(d\phi^2+e^{2\phi}d\bar\ga d\ga),\quad
H_1=-\jp e^{2\phi}d\phi\w d\bar\ga\w d\ga\eqno(17)$$
and the $D$-brane $P_1$ is characterized by (27). It remains to
determine $\al_1$.
 We use the formula
(13) and argue that the second term on its  r.h.s.
 vanishes in this
 particular case.
Indeed, we can choose the map $g_1$ equal to $g_1(f_1)=1$.
Hence we conclude
$$\al_1 =\js e^{2\phi}d\ga\w d\bar\ga.\eqno(28)$$
It is important to note that the variables $\phi$ and $\ga$ in (28)
are subject to the constraint (27).

Having obtained the quintuple $(T_1,d\si_1^2,H_1,P_1,\al_1)$,
we can write
down the action for open strings.  In the spirit of
the discussion in Sec 2.1,
we pick up some globally defined two-form potential $B_1$
 such that $dB_1=H_1$.
We know that the  combination
$B_1-\al_1$ on the $D$-brane is always a closed form. In the
 present case
it is even exact therefore
it has a globally defined
potential $A_1$ on $P_1$. The non-metric part of the $\si$-model
action (22) on the upper half-plane  can now be written as follows
$$i\int_{S_\da}B_1+i\int_{\vert z\vert =1} A_1.$$ Actually,
in our situation, we can take
$$B_1=\js e^{2\phi}d\ga\w d\bar\ga. $$
Clearly, the difference $B_1-\al_1$ on the $D$-brane $P_1$ vanishes
hence   $A_1=0$. In other words, we can cast the open string
action without any boundary term as follows
$$S=-{i\over 2}\int_{S_\da} d\bz\w dz (\d_z\phi \d_\bz \phi
+e^{2\phi}\d_z\ga \d_\bz\bar\ga).\eqno(29)$$
We stress again that this is the {\it open} string action.
Nevertheless
the integration over the boundary $\az=1$ is missing due to the
 clever choice
of the potential $B_1$. The boundary conditions that accompany
the action (29)
say that the boundary of the world-sheet lies in the $D$-brane $P_1$.

We could be happy that in the particular case of $AdS_2$ branes
in the Euclidean $AdS_3$
the open string action can be written without any extensions
of the type
$x_p$ and $\ti x$ (cf. (2)).
  On the other hand, the action
(29) is expressed in the coordinates $\phi,\ga$ and this fact
makes technically complicated to verify whether there is a
a residual loop group symmetry. Indeed, the loop group action on the
configuration $\phi,\ga$ is given by a cumbersome formula and it
would be  tedious to check the residual symmetry by working directly
with (29).
It turns out that it is much easier to treat the
symmetry issues  in
the $(\int_\Om \ti x^*H-\int_{S^\ua}x_P^*\al)$-representation
of the non-metric part of the open string action, because
it can be done
without the necessity of introducing any coordinates on the
target $SL(2,\bc)/SU(2)$.  Let us see how this works.

The points in the coset  $T_1=SL(2,\bc)/SU(2)$ are Hermitian
$2d$-matrices $h$
with unit determinant and positive trace. We wish to express the
data $d\si_1^2, H_1,P_1,\al_1$ in terms of $h$. Here is the answer
$$d\si_1^2 =-{1\over 8}{\rm Tr} (dhh^{-1} dh h^{-1});\eqno(20a)$$
$$H_1=-{1\over 24}{\rm Tr}(dhh^{-1}\w [dhh^{-1}\st dhh^{-1}]);
\eqno(20b)$$
$$P_1=\{h ; {\rm Tr}(h\si_1)=c\};\eqno(30a)$$
$$\al_1= +{c\over 16 +4c^2}{\rm Tr}(\si_1 h\si_1 dh\w \si_1 dh),
 \quad h\in P_1.\eqno(30b)$$
Here we recall that $\si_1=\left(\matrix{0&1\cr 1&0}\right)$
is the
Pauli matrix.
The reader can easily verify, that inserting the parametrization
(25) into the formulae (20ab) and (30ab), gives respectively
formulae
(17) and (27,28).

Thus we conclude the discussion of the $AdS_2$
branes in  the $SL(2,\bc)/SU(2)$ background by writing for them the
 open string action
in the coordinate free way:
$$S=-{i\over 4}\int_{S_\da}
 d\bz\w dz {\rm Tr}(\d_\bz hh^{-1}\d_z hh^{-1}) $$
$$-{i\over 24}\int_\Om {\rm Tr}(d\ti h\ti h^{-1}\w
[\ti h\ti h^{-1}\st
 d\ti h\ti h^{-1}])
 -i{c\over 16 +4c^2}\int_{S^\ua}{\rm Tr}(\si_1 h_P\si_1
dh_P\w \si_1 dh_P)
.\eqno(31)$$
Recall once more that $h_P$ (lying
in the brane $P_1$) is the $S^\ua$-extension of
(the open string
configuration)  $h$  and $\ti h$ is the extension of $h\cup h_P$
to the
interior of the ball $\Om$. We shall show in   section 4, that the
action (31)
 possesses a residual $SL^a(2,\br)$ loop group symmetry.

\vskip1pc

\subsubsection{The target $D/G_2=\slc/SL^a(2,\br)$}
In order to find the shape
of the dual $D$-branes, it is convenient
to switch from the parametrization (15) of the coset
$\slc/SL^a(2,\br)$
to the parametrization by the  Hermitian 2$d$-matrices with
determinant equal
 to  $(-1)$.
They can be written as
$$s=\left(\matrix{u&w\cr\bar w &v}\right),
 \quad uv-\bar ww=-1.\eqno(32)$$
Clearly, $\slc/SL^a(2,\br)$ is nothing but the de Sitter space,
if it is equipped with the
Minkowski metric
$$ds_{dS}^2=dudv-d\bar wdw $$
restricted to the surface $uv-\bar ww=-1$. Two parametrizations
are related
by $s=f_2\si_1f_2^\dagger$ (cf. (15)), that gives
$$\jp(u+v)=L,\quad \jp(u-v)=L\cos{2\chi}+\sin{2\chi}\cos{2\vth},
\eqno(33a)$$
$$w=\cos{2\chi}-L\sin{2\chi}
\cos{2\vth}-i\od\sin{2\chi\sin{2\vth}}.\eqno(33b)$$
The reader may verify that, indeed, it holds $\quad uv-\bar ww=-1$.

\vskip1Pc
\noindent The coset projection $s:SL(2,\bc)\to \slc/SL^a(2,\br)$
is then given by
$$s(l) =l\si_1l^\dagger.\eqno(34)$$
The shape of the dual de Sitter $D$-brane is now given by the
projection of the first-order $D$-brane (24):
$$s(p)=\left(\matrix{\nu^2c&1+i\nu\lm c\cr 1-i\nu\lm c& \lm^2 c}
\right).$$
If $c=0$, then the dual $D$-brane is just a point $w=1$, $u=v=0$.
If $c\neq 0$, then the dual $D$-brane $P_2$
is two-dimensional and is characterized
by the following relations
$${\rm Re}(w)=1,\quad   {\rm sign}(u) ={\rm sign}(v)=
{\rm sign}(c).\eqno(35)$$
Thus the dual $AdS_2$ brane has a conical shape.
 The remaining missing ingredient of the open string quintuple
is the form $\al_2$ on $P_2$.
It is obtained from (13) but the straightforward calculation
based on the
coset parametrization (15) would be exceedingly tedious. Instead,
we shall use
the fact that the form $\al_2$ is defined only on the brane $P_2$.
In this case,
 we can parametrize the section   $f_2$ as follows
$$f_{P_2}=\left(\matrix{1+{i\over 2}\sqrt{uv}&\jp u\cr
\jp v &1-{i\over 2}\sqrt{uv}}\right),\eqno(36)$$
since it is easy to see that the coset projection map $s$ gives
$$s(f_{P_2})=f_{P_2}\si_1f_{P_2}^\dagger=
\left(\matrix{u&1+ i\sqrt{uv}\cr
 1-i\sqrt{uv}&v}\right).\eqno(37)$$
This is indeed the $D$-brane surface $P_2$ because the matrix on
 the r.h.s. of
(37) is the general solution of the relations (35).

Using the parametrization (36), it is now easy to calculate the
 form
$\al_2$ from the formula (13). The result is simple:
$$\al_2={i\over 4}{du\w dv\over \sqrt{uv}}.\eqno(38)$$
 Of course, for $c=0$ the brane reduces to a point and the
form $\al_2$
automatically vanishes.

Thus we conclude the discussion of the de Sitter dual of the $AdS_2$
branes  by writing for them the
 open string action
in the  global   way (3):
$$S=-{i\over 4}\int_{S_\da}
 d\bz\w dz {\rm Tr}(\d_\bz ss^{-1}\d_z ss^{-1}) $$
$$-{i\over 24}\int_\Om {\rm Tr}(d\ti s\ti s^{-1}
\w[d\ti s\ti s^{-1}\st
 \d\ti s\ti s^{-1}])+
 {1\over 4}\int_{S^\ua} {du_P\w dv_P\over \sqrt{u_Pv_P}}
.\eqno(39)$$
Recall once more that $s_P=\left(\matrix{u_P&1+ i\sqrt{u_Pv_P}\cr
 1-i\sqrt{u_Pv_P}&v_P}\right)$ is the $S^\ua$-extension of $s$ to
 the brane $P_2$ and $\ti s$ is the extension of $s\cup s_P$ to the
interior of the ball $\Om$. We shall show in  Section 4, that the
action (39) possesses a residual $SL^a(2,\br)$ loop group symmetry.

\subsection{Spherical branes}
 We continue with the case $M=SU(2)$ embedded
into $\slc$ in the standard way:
$$\left(\matrix{\al &-\bar\bt\cr\bt &\bar \al}\right),\quad
\al,\bt\in{\bf C},\quad
\al\bar\al+\bt\bar\bt=1.$$
The possible first order $D$-branes are then the submanifolds $Md$ of
 $SL(2,\bc)$
with $d$ being a fixed element of $SL(2,\bc)$. As in the previous
 case of the $AdS_2$ branes,
we do not make here the most
general choice of $d$; instead, we consider only $d$ having form
$$d=\left(\matrix{e^\rho& 0\cr 0 &e^{-\rho}}\right),\quad
 \rho\in\br.$$
The motivation for this particular choice is simple:
if we switch from the first order action (10) to the
second order one (22), it  reproduces
the $spherical$ branes considered in $\cite{PST}$.
\vskip1pc
\noindent Thus our first order $D$-branes are three-dimensional
submanifolds
of $SL(2,\bc)$
of the form
$$p=\left(\matrix{e^\rho\al &-e^{-\rho}\bar\bt\cr e^{\rho}\bt &
e^{-\rho}\bar \al }\right),\eqno(40)$$
where the real number $\rho$ is fixed and
$\al,\bt\in\bc$ vary while respecting
the constraint $ \al\bar\al+\bt\bar\bt=1$.
\subsubsection{The target $D/G_1=SL(2,\bc)/SU(2)$.}
Now we want to find the shape of the  corresponding $D$-brane $P_1$
in the target $D/G_1=SL(2,\bc)/SU(2)$.
 It is given by the coset projection (26) of the first order
 $D$-brane (40):
 $$h(p)=\left(\matrix{{\rm cosh}2\rho-{\rm sinh}2\rho(\al\bar\al
 -\bt\bar\bt) &
  -2\al\bar\bt {\rm sinh}2\rho
\cr   -2\bar\al\bt {\rm sinh}2\rho &
 {\rm cosh}2\rho+{\rm sinh}2\rho(\al\bar\al -\bt\bar\bt) }\right).$$
Note that the trace of $h(p)$ is constant.
Comparing with the parametrization (25), we thus observe that the
  $D$-brane
$P$ in the coset $\slc/SU(2)$ is characterized by the equation
$${\rm Tr} (h)=
e^\phi(1+\bar\ga \ga)+e^{-\phi}=2{\rm cosh}2\rho\eqno(41)$$
with constant $\rho$. This is exactly the spherical  brane
 considered in
\cite{PST}.

As we have said, the first order action (10) together with the
 choice of the
first order $D$-brane (40) (and $\al_D=0$)
determines the open string quintuple
$(T_1,d\si_1^2,H_1,P_1,\al_1)$.
 So far we have determined  its four elements: the target
$T_1=\slc/SU(2)$ is parametrized as in (25), then
$$ds_1^2=  -{1\over 8}{\rm Tr} (dhh^{-1} dh h^{-1}) =
 -\js(d\phi^2+e^{2\phi}d\bar\ga d\ga), $$
$$  H_1=-{1\over 24}{\rm Tr}(dhh^{-1}\w [dhh^{-1}\st dhh^{-1}])=
-\jp e^{2\phi}d\phi\w d\bar\ga\w d\ga$$
and the spherical $D$-brane $P_1$ is characterized by (41).

It remains to determine $\al_1$.
Of course, we use the general formula (13). The calculation
 is a bit tedious
 but straigthforward
and it gives a remarkably simple result:
 $$\al_1 = -{{\rm cosh}2\rho\over 8{\rm sinh}^22\rho}{\rm Tr}
(hdh\w dh),
\quad {\rm Tr}(h)=2{\rm cosh}2\rho.
\eqno(42)$$
 It is crucial to stress that  the form $\al_1$ in the formula (42)
is defined only on the surface (41).

Having obtained the quintuple $(T_1,ds_1^2,H_1,P_1,\al_1)$, we
 can write
down the action for open strings attached
to the spherical brane. The result is
clearly
$$S=-{i\over 4}\int_{S_\da}
 d\bz\w dz {\rm Tr}(\d_\bz hh^{-1}\d_z hh^{-1}) $$
$$-{i\over 24}\int_\Om {\rm Tr}(d\ti h\ti h^{-1}
\w[\ti h\ti h^{-1}\st
 d\ti h\ti h^{-1}])
+ i{{\rm cosh}2\rho\over 8{\rm sinh}^22\rho}\int_{S^\ua}{\rm Tr}
( h_P dh_P\w  dh_P)
.\eqno(43)$$
Recall once more that $h_P$ is the $S^\ua$-extension of $h$ lying
in the brane $P_1$ and $\ti h$ is the extension of $h\cup h_P$ to the
interior of the ball $\Om$. The boundary conditions that
accompany the
 action (43)
say that the boundary of the world-sheet lies in the $D$-brane (41).
We shall show in Section 4, that the
action (43)
possesses a residual $SU(2)$ loop group symmetry.

\subsubsection{The target $D/G_2=\slc/SL^a(2,\br)$}
It is straigthforward to find the shapes of the de Sitter duals
of the spherical $D$-branes. We just apply the coset projection
 map (34)
on the first order brane (40) and obtain
$$s(p)= \left(\matrix{-\al\beta -\bar\al\bar\beta&\al^2-
\bar\beta^2\cr
\bar\al^2-\beta^2&
+\al\beta +\bar\al\bar\beta}\right).$$
Note that the resulting brane $P_2$
  can be characterized  in
the coordinate free way as
$$P_2=\{s;{\rm Tr}(s)=0.\}$$
In the parametrization (32), this means $u+v=0$. Remembering that
det$s=-1$ this gives $u^2+w\bar w=1$ which is the equation of the
sphere. Thus the dual $D$-branes of the spherical branes are also
spherical.

The form $\al_2$ is given by the formula (13). We work in the
coordinates
$\chi,\vartheta,L$ in which the  $D$-brane $P_2$ is characterized by
the equation $L=0$. It turns out that for $g_2(f_2)$ appearing in
(13) we can take a constant map
$g_2(f_2)=\left(\matrix{e^\rho& 0\cr 0 &e^{-\rho}}\right)$, hence
the second term in (13) vanishes. A direct calculation
then shows that, on the surface $L=0$, the first term also vanishes.
Thus we obtain
$$\al_2=0$$
 We conclude the discussion of the de Sitter dual of the spherical
branes  by writing for them the
 open string action:
$$S=-{i\over 4}\int_{S_\da}
 d\bz\w dz {\rm Tr}(\d_\bz ss^{-1}\d_z ss^{-1}) -
{i\over 24}\int_\Om  {\rm Tr}(d\ti s\ti s^{-1}\w[\ti s\ti s^{-1}\st
 d\ti s\ti s^{-1}])
  .\eqno(44)$$
Recall once more that $s_P$ is the $S^\ua$-extension of $s$ to
 the brane $P_2$ and $\ti s$ is the extension of $s\cup s_P$ to the
interior of the ball $\Om$. We shall show in   Section 4, that the
action (44) possesses a residual $SU(2)$ loop group symmetry.

\subsection{Euclidean $AdS_3$ branes}

This case was not discussed in \cite{PST}. The reason is simple: the
leading thread of our reasoning
 in this section is T-duality, while Ponsot, Schomerus and Teschner
have organized their  paper from the point of view of the
residual symmetry
of the $D$-brane configurations. They have considered only the case
were this   symmetry was semi-simple. However, the Euclidean
$AdS_3$ branes,
that we are going to study now, have a solvable residual symmetry.
 Consider thus  the remaining case $M=AN$ embedded into $SL(2,\bc)$ in
the
standard way (cf. (16) ):
$$b=\left(\matrix{e^{\phi/2}&0\cr \ga e^{\phi/ 2}&e^{-{\phi/2}}}
\right),
\quad \phi\in{\bf R}, \ga\in\bc\eqno(45)$$
 The possible first order $D$-branes are then the submanifolds $Md$
of $SL(2,\bc)$
with $d$ being a fixed element of $SL(2,\bc)$. As usual, we do not
make here the most
general choice of $d$; instead, we consider only the simplest
 possible $d$ having the form
$$d=\left(\matrix{1 & 0\cr 0 &1}\right) .$$
 Thus our first order $D$-branes are three-dimensional submanifolds
 of $SL(2,\bc)$
of the form $p=b$ (cf. (45)), where $\phi\in\br$ and $\ga\in\bc$
  vary.
 \subsubsection{The target $D/G_1=SL(2,\bc)/SU(2)$.}

\noindent  Now we are ready to find the shape of the $D$-brane
$P_1$ in $D/G_1$.
It is given by the coset projection (26) of the first order
 $D$-brane (45):
 $$h(p)\equiv h(b)=\left(\matrix{e^\phi &  e^\phi \bar \ga\cr
 e^\phi\gamma &e^\phi\ga\bar\ga
+e^{-\phi}}\right), \quad  \phi\in\br,\  \ga\in\bc.\eqno(25)$$
Thus
 we immediately observe that the $D$-brane
$P_1$ coincides with the whole  target $\slc/SU(2)$.   This is the
reason why
 we  call
$P_1$ the Euclidean $AdS_3$ brane. In fact, whatever $d$ we choose,
 it turns
out that the corresponding $D$-brane $P_1$ sweeps the whole target
 space
 $\slc/SU(2)$.
However, the choice of $d$ has an influence on the shape of the dual
$D$-brane $P_2$ in $\slc/SL(2,\br)$.

As we have said, the first order action (10) together with the
 choice of the
first order $D$-brane (45) (and $\al_D=0$)
determines the open string quintuple
$(T_1,ds_1^2,H_1,P_1,\al_1)$. So far we have determined  its four
elements:
the target
$T_1=\slc/SU(2)$ is parametrized as in (25), then
$$d\si_1^2=-\js (d\phi^2+e^{2\phi}d\bar\ga d\ga),\quad  H_1=-\jp
e^{2\phi}d\phi\w d\bar\ga\w d\ga$$
and the $D$-brane $P_1$ coincides with the target $T_1$.
It remains to determine $\al_1$. We use the formula
(13) and argue that the second term on its r.h.s. vanishes in this
 particular case.
Indeed, we can choose the map $g_1(f_j)$ equal to $g_1(f_j)=1$.
Hence we conclude
$$\al_1 =\js e^{2\phi}d\ga\w d\bar\ga.$$
The open string
action  therefore reads
$$S=-{i\over 2}\int_{S_\da} d\bz\w dz (\d_z\phi \d_\bz \phi
+e^{2\phi}\d_z\ga \d_\bz\bar\ga).
 \eqno(46)$$
  The boundary conditions that accompany the action (46)
say that the boundary of the world-sheet is not constrained but
can be located everywhere on the $AdS_3$ target.
We shall show in Section 4, that the action (46) possesses
a residual $AN$ current symmetry.
\subsubsection{The target $D/G_2=\slc/SL^a(2,\br)$}
The derivation of the T-dual quintuple
 $( T_2,d\si_2^2, H_2, P_2,\al_2)$
is  straightforward. We have to apply the coset projection map
(34) to the first-order brane (45). We obtain
$$s(p)\equiv s(b)=\left(\matrix{0&1\cr 1&\ga+\bar\ga}\right).$$
In the parametrization (32), the dual de Sitter brane $P_2$ is
therefore just a
line $w=1,u=0$. This means, in particular, that the two-form
$\al_2$
on it automatically vanishes.

Thus we conclude the discussion of the de Sitter dual of the
Euclidean
$AdS_3$
branes  by writing for them the
 open string action:
$$S=-{i\over 4}\int_{S_\da}
 d\bz\w dz {\rm Tr}(\d_\bz ss^{-1}\d_z ss^{-1})
-{i\over 24}\int_\Om
 {\rm Tr}(d\ti s\ti s^{-1}\w[d\ti s\ti s^{-1}\st
 d\ti s\ti s^{-1}]).$$
Recall once more that $s_P$ is the $S^\ua$-extension of $s$ lying
in the one-dimensional
brane $P_2$ and $\ti s$ is the extension of $s\cup s_P$ to the
interior of the ball $\Om$. We shall show in Section 4, that this
action   possesses a residual $AN$ loop group symmetry.

\section{$D$-branes and the residual   loop  symmetry}
 In   Section 3,
we have described three types of branes in
the $SL(2,\bc)/SU(2)$ WZW model
and in its de Sitter dual. We have
 referred to them, respectively, as
$AdS_2$, spherical and Euclidean $AdS_3$ branes. We shall now argue
that each type
has a residual loop group
 symmetry  respecting the $D$-branes boundary conditions
$(P,\al)$.
We say {\it residual}, because there is even
bigger loop group  symmetry of the bulk model. We shall first
 describe it
and then discuss the symmetry of various $D$-brane configurations.
\subsection{Bulk loop group symmetry}
Consider  antiholomorphic maps $g(\bz)$ from the Riemann sphere
 without poles
into the complex group $SL(2,\bc)$. The set of such maps
form a loop group $LSL(2,\bc)$. The $LSL(2,\bc)$ loop group
symmetry of
 the first order bulk action
$$S=\jp\int_{\d\Om} d\si\w  d\tau {\rm ImTr}
(\d_\tau ll^{-1}\d_\si ll^{-1})
+{1\over 12}\int_\Om{\rm ImTr}(dll^{-1}\w [dll^{-1}\st dll^{-1}])$$
$$-\jp\int_{\d\Om} d\si \w d\tau{\rm ImTr} (\d_\si ll^{-1}
 i\d_\si ll^{-1}).
\eqno(14)$$
  is the direct consequence of the Polyakov-Wiegmann formula (11).
Indeed, it  is the matter of an easy check to see that the bulk
action (14)
does not change its value upon the replacing a configuration
$l(z,\bz)$ by $g(\bz)l(z,\bz)$. In particular, also the field
equations following from (14)
$$l^{-1}\d_zl=0$$
are manifestly $LSL(2,\bc)$-symmetric since a solution $l(\bz)$
becomes  clearly another solution upon the
transformation
$$l(\bz)\to g(\bz)l(\bz),\quad g(\bz)\in LSL(2,\bc).\eqno(47)$$
The $LSL(2,\bc)$ symmetry transformation shows up also in the
second order formalism. For concreteness, consider the target
$SL(2,\bc)/SU(2)$ and the bulk $\si$-model (22)
$$S=-{i\over 4}\int_{\d\Om}
 d\bz\w dz {\rm Tr}(\d_\bz hh^{-1}\d_z hh^{-1})
-{i\over 24}\int_\Om
{\rm Tr}(d\ti h\ti h^{-1}\w[d\ti h\ti h^{-1}\st
 d\ti h\ti h^{-1}]).\eqno(22)$$
 Recall once more that   $\ti h$ is the extension of $h$ to the
interior of the ball $\Om$.  First note that from  the first
order configuration $l$ we obtain the second order trajectory $f_1$
by the (Iwasawa)  decomposition (6). This means (cf. (19)) that
$$h=f_1f_1^\dagger=ll^\dagger\eqno(19b)$$
and the transformation (47) translates into
$$h(z,\bz)\to g(\bz) h(z,\bz) g^\dagger(z).\eqno(48)$$
Then note that the Polyakov-Wiegmann formula (11) holds note
only for the indefinite metric $(.,.)_\D$ but also for every
invariant bilinear form on $Lie(D)$, in particular for Tr appearing
in (22). Using this fact, it is the matter of an easy calculation
 to see
that the action (22) is indeed invariant with respect to the
loop group transformation (48).

In the de Sitter dual, the bulk action
$$S=-{i\over 4}\int_{S_\da}
 d\bz\w dz {\rm Tr}(\d_\bz ss^{-1}\d_z ss^{-1})
-{i\over 24}\int_\Om
{\rm Tr}(d\ti s\ti s^{-1}\w[d\ti s\ti s^{-1}\st
 d\ti s\ti s^{-1}])
  \eqno(23)$$
has the symmetry
$$s(z,\bz)\to g(\bz) s(z,\bz) g^\dagger(z)$$ which originates
from the first order transformation (47) by the coset projection map
$s=l\si_1l^\dagger$ (cf. (34)).
\subsection{Residual loop group symmetry}
\subsubsection{First order formalism}
Consider the first order brane $\Pi=Md$ in the double $D=SL(2,\bc)$,
 where $M$ is some maximally
isotropic subgroup of $D$. We know that the open string
first-order action reads (cf. (10))
$$S[l]=\jp\int_{S_\da} d\si\w  d\tau
{\rm ImTr}(\d_\tau ll^{-1}\d_\si ll^{-1})
-\jp\int_{S_\da} d\si \w d\tau
{\rm ImTr}(\d_\si ll^{-1}i\d_\si ll^{-1})$$
$$+{1\over 12}\int_\Om{\rm ImTr}
(d\ti l\ti l^{-1}\st [d \ti l \ti l^{-1}\st
 d\ti l\ti l^{-1}]),\eqno(49)$$
where $l$ is the true open string configuration defined on $S_\da$
whose boundary $l(\az=1)$ takes values in $Md\subset D$,
  $l_{\Pi}$ is
its $S^\ua$ extension into the first order brane $\Pi=Md$ and
 $\ti l$
is the extension of   $l\cup l_{\Pi}$ to the ball $\Om$.

The open string action (49) cannot be symmetric with respect to all
 $LSL(2,\bc)$ loop transformations $l(z,\bz)\to g(\bz) l(z,\bz)$
( $\az\leq 1$). Indeed, an arbitrary
$LSL(2,\bc)$ element $g(\bz)$ do not preserve the first-order brane
 boundary condition
saying that $l(\az=1)\in Md$. On the other hand, we can consider
the residual subgroup $LM_0\subset LSL(2,\bc)$ consisting of
the elements
$g(\bz)$ satisfying $g(\az=1)\in M$ and having
the property that the induced mapping from the equator of the
 Riemann
sphere into the group $M$ is homotopically trivial
(of course, this is always
true if the group $M$ is simply connected).
 The action $l\to ml$ of this
residual subgroup transforms one open string configuration
$l(z,\bz)$
into another one $m(\bz)l(z,\bz)$ while respecting the boundary
conditions $(ml)(\az=1)\in Md$.
 It makes therefore sense to ask whether
the action (49) changes upon such a transformation.

In order to answer this question, we must extend the transformed
configuration $ml$ from $S_\da$ to
$S^\ua$ in such a way that the $S^\ua$ piece
lies in the first order brane $\Pi=Md$. This is easy,
we first consider the original extension $l_{\Pi}$ of $l$
and then any map
$m_{\Pi}:S^\ua\to M$ satisfying $m_{\Pi}(\az=1)=m(\az=1)$,
where $m(\bz)\in LM_0$ (the map $m_\Pi$ always exists due to our
assumption about the homotopical triviality). Thus we set
$(ml)_\Pi=m_\Pi l_\Pi$.
In the analogous way, the tilde-extension of the transformed
configuration $ml$ can be written as $\widetilde{(ml)}=
\ti m\ti l$ for
 an appropriate
$\ti m$. Now we replace $l$, $l_{\Pi}$ and $\ti l$
in (49) by $ml$, $m_{\Pi}l_{\Pi}$ and $\ti m\ti l$ and use the
Polyakov-Wiegman formula (11).  The result is immediate:
$$S[ml]=S[l]-\jp\int_{S^{\ua}}{\rm ImTr}( m^{-1}_{\Pi}dm_\Pi
\w dl_\Pi l_\Pi^{-1}).\eqno(50)$$
Let us argue that the second term in the r.h.s. of (50)
vanishes therefore the action (49) is  invariant
with respect to the residual loop group symmetry $LM_0$. Indeed,
this follows from the isotropy of the $Lie(M)$ with respect
to the ImTr and from the fact that both forms
 $m^{-1}_{\Pi}dm_\Pi$ and
$dl_\Pi l_\Pi^{-1}$ are $Lie(M)$-valued.
\subsubsection{Second order formalism}
In principle, the second order formalism follows from the
first order one and, therefore, we can consider that we have
 already
proven in the precedent subsection 4.2.1
the residual loop group symmetry of all $D$-brane
 configurations considered
in this paper.   On the other hand, the reader might
wish to see the
proof of the residual symmetry by working directly in
the second order
formalism. In three cases out of six, we do not have
 a coordinate
free expression for the two-form $\al$ on the brane.   Due to
 this circumstance, we have to work in coordinates,
 the relevant formulae for the loop
group action
 are quite complicated and
we do not detail the second-order symmetry demonstration here.
   However, in the three remaining
cases (including all brane configurations considered by
Ponsot, Schomerus and
Teschner \cite{PST}) we do have the coordinate free expression
for the form $\al$ and this fact makes possible to give
the simple  second-order  proof of the residual loop symmetry.
As an example, we prove the residual $SU(2)$ loop symmetry
of the spherical  brane in the $SL(2,\bc)/SU(2)$ background.
 Recall
that its action principle reads
$$S=-{i\over 4}\int_{S_\da}
 d\bz\w dz {\rm Tr}(\d_\bz hh^{-1}\d_z hh^{-1}) $$
$$-{i\over 24}\int_\Om {\rm Tr}(d\ti h\ti h^{-1}
\w[d\ti h\ti h^{-1}\st
 d\ti h\ti h^{-1}])
 + i{{\rm cosh}2\rho\over 8{\rm sinh}^22\rho}
\int_{S^\ua}{\rm Tr}( h_P dh_P\w  dh_P)
.\eqno(43)$$
Recall once more that $h_P$ is the $S^\ua$-extension of $h$ lying
in the brane $P_1$ and $\ti h$ is the extension of
$h\cup h_P$ to the
interior of the ball $\Om$. The boundary conditions that
 accompany the
 action (43)
say that the boundary of the world-sheet lies in the $D$-brane (41).

We know from the first order analysis that the residual symmetry
group is $LSU(2)$, i.e. the group of maps $g(\bz)$ such that
$g(\az=1)\in SU(2)$. We shall denote the elements of $LSU(2)$
as $m(\bz)$ and, as in the precedent subsection, to every $m(\bz)$
we associate also the configurations $m_\Pi$ and $\ti m$.
We remark that $m_\Pi^\dagger=m_\Pi^{-1}$ and we use the
Polyakov-Wiegmann
formula (11) to
calculate the transformed action
$$S[mhm^\dagger]= S[h]+$$
$$+{i\over 4}\int_{S\ua}{\rm Tr}\{m_\Pi^{-1}dm_\Pi
\w(h_P^{-1}dh_P +dh_P h_P^{-1})-m_\Pi^{-1}dm_\Pi \w h_P
m_\Pi^{-1}dm_\Pi h_P^{-1}\}+$$
$$+i{{\rm cosh}2\rho\over 8{\rm sinh}^22\rho}\int_{S^\ua}
{\rm Tr}(m_\Pi h_pm_\Pi^{-1} d(m_\Pi h_pm_\Pi^{-1})\w
(m_\Pi h_pm_\Pi^{-1})).$$
At a first sight, it is not evident that the contributions  from
 the second and third
line cancel each other. However, it is indeed true due to the
identity
$$h_P^{-1}=2{\rm cosh}2\rho -h_P\eqno(51)$$
holding on the surface of the $D$-brane
$P_1=\{h;{\rm Tr}(h)=2{\rm cosh}2\rho\}$.
\vskip1pc
\noindent The identity (51) has its cousins for all other branes
for which  the form $\al$ can be written in the coordinate free way.
We list them in order to facitilitate the work of the reader
who wishes
to check the second order residual symmetry also for them. It is
$$(h\si_1)^{-1}=h\si_1 -c, \quad {\rm Tr}(h\si_1)=c,$$
for the $AdS_2$ branes in the $\scu$ model and
$$s^{-1}=s,\quad {\rm Tr}(s)=0$$
for the spherical branes in the de Sitter background.
We recall that $s$ and $h$ are Hermitean matrices, det$h$=1,
det$s=-1$
and Tr$h$ is positive.
\section{Conclusions and outlook}
The possibility  to formulate the $\scu$ model in the
 first order way (14)
was a fruit of the Poisson-Lie symmetry of this theory \cite{K}.
 This
property then leads to the T-dualizability of the model with the
 T-dual
being the de Sitter space. In this paper, we have applied the
 general
theory \cite{KS3,KS4} of the $D$-branes Poisson-Lie T-duality
to this particular
example.  We stress, however, that our results may find applications
also in the direct study of the $\scu$ model without concentrating
on the T-duality story. In particular, we have been able to
formulate
the dynamics of the $AdS_2$ brane  in the coordinate independent
 manner (31)
which could  render more transparent the geometrical structures
involved
in this system.
 As it is well-known \cite{LorAdS},  several
technical issues concerning  string propagation in the Lorentzian
$AdS_3$ can be Wick-continued to the Euclidean case treated here.
It is an interesting open question how the T-duality  discovered
in the Euclidean context manifests itself in the Lorentzian one.

\end{document}